# An Efficient Routing Protocol in Mobile Ad-hoc Networks by Using Artificial Immune System

Fatemeh Sarkohaki
Department of Computer Engineering
Germi branch, Islamic Azad
University, Germi, Iran
Ahf_1365@yahoo.com

Reza Fotohi*
Department of Computer Engineering
Germi branch, Islamic Azad
University, Germi, Iran
fotohi@ieee.org

Vahab Ashrafian
Department of Computer, Science and Research
Branch, Islamic Azad University, Ardabil, Iran
vahabashrafi@gmail.com

*Abstract*— Characteristics of the mobile ad-hoc networks such as nodes high mobility and limited energy are regarded as the routing challenges in these networks. OLSR protocol is one of the routing protocols in mobile ad hoc network that selects the shortest route between source and destination through Dijkstra's algorithm. However, OLSR suffers from a major problem. It does not consider parameters such as nodes' energy level and links length in its route processing. This paper employs the artificial immune system (AIS) to enhance efficiency of OLSR routing protocol. The proposed algorithm, called AIS-OLSR, considers hop count, remaining energy in the intermediate nodes, and distance among node, which is realized by negative selection and ClonalG algorithms of AIS. Widespread packet - level simulation in ns-2 environment, shows that AIS-OLSR outperforms OLSR and EA-OLSR in terms of packet delivery ratio, throughput, end-end delay and lifetime.

## I. INTRODUCTION

MANET is a mobile ad hoc network, temporary and instantaneous networks that develops for special purpose. Indeed, wireless networks are collection of wireless mobile nodes which are infrastructure less, autonomous and without any centralized management networks. Therefore, nodes in this type of network are responsible for dynamically discovering each other. Based on nature of dynamic, the network topology of this type of network change continuously. Because manet are mobile, connections changing are unpredictable. The biggest challenge of this kind of networks are faced with, routing packet efficient till reach to destination without creation overhead. So, must be proposed some methods for routing that can route with overhead less. Several routing algorithms are presented by MANET networks, which each of them have features, advantages and disadvantages.

There are various methods of classifying routing protocols in mobile ad-hoc networks; however, most of them depend on routing strategy and network structure. In general, there are two types of routing protocols: first, is table-driven or proactive routing in which protocols try to get comprehensive, updated information of each node of network. In other word, these protocols save routes' information even they are not using. Therefore, each node requires one or more tables to maintain routing information. The second type is on demand or reactive. These types of protocols create and find a route in terms of supply with overflow transferring of request packets, once source tries to send a message. [1]. Optimized link state routing (OLSR) protocol is a table-driven routing protocol in mobile ad hoc network routing [2], discussed in many studies. OLSR protocol works based on Dijkstra's algorithm which, in turn, determines the shortest (but not necessarily most accurate) route based on hop counts. The shortest route might have a larger delay or its nodes might have congestion and, then, the data packets are dropped once reaching to them. High speed of some nodes in short routes might also lead to a sooner failure of the routes. Therefore, route selection in this protocol is controlled by a large number of variables [3]. In this work, an attempt is made to improve OLSR protocol using artificial immune system for optimum routing of the mobile ad hoc networks. To improve routing process, parameters including remaining energy in the route intermediate nodes, hop counts, and distance between the intermediate nodes have been applied. This paper is organized as follows: In second Section, OLSR protocol in mobile ad hoc networks is introduced. Section 3 introduces the artificial immune system. Section 4 is allocated to introducing works carried out on artificial immune system. In section 5, the proposed method is discussed. Section 6 evaluated the efficiency of the proposed method. Finally, section 7 brings the concluding remarks.

## II. OLSR ROUTING PROTOCOL

As a proactive protocol, OLSR is a routing protocol presented by mobile ad hoc networking (MANET) in the internet engineering task force (IETF) for mobile ad hoc network [4, 5 and 6]. The network nodes alternatively exchange topology information to each other, so the optimum route between two nodes is constantly available. OLSR is also a link state protocol. The difference between the optimization performed in this protocol as compared to that of other link establishes in the creation of MPR concept. Within this protocol, the network nodes are required to select a bunch of their neighbors as the MPR group. The group is needed to be selected in a way that all nodes have a two hop distance with their selector node. A given node (for example node N) which is selected as the MPR node, alternatively transmits the information to network from its selector node. These alternative messages are delivered and processed by all neighbors of the node N, but only MPR neighbors of node N resend them. Indeed, this mechanism not only reduces the network control overload, but also introduces a limited number of links to the network nodes [7, 8, 9 and 10]. As the first step, OLSR recognizes its neighbors through sending Hello packets to the neighbors around each node. Then, using the information obtained, it creates a table indicating the relationship between the nodes with the neighbors. Next, the nodes will transmit their information with their number in a TC packet to the neighboring nodes. However, TC packets transmission is performed using the MPR nodes. In this way, all nodes presented in the network are aware about the existing information and their connection with other nodes. This information are stored in a table for each node. As the next step, each node must select the optimum route for the neighboring nodes using the collected information. The route selection process is carried out based on the least hop counts through Dijkstra's algorithm. After this step, each node is provided with a routing table containing the optimum routes to neighboring nodes. In this case, network is stable [11, 12, 13, 14 and 15]. Once switching nodes location, the abovementioned process is repeated and the tables are updated.

### III. ARTIFICIAL IMMUNE SYSTEM

The artificial immune systems are designed based on the available knowledge functions of the immune system in vertebrates. Generally, the artificial immune systems are algorithms inspired by biology. These are computer algorithms where their principles and characteristics are defined based upon studying the adaptive properties a, resistance of the biological samples as well. The artificial immune system is a pattern of machine learning .Machine learning is the computer ability to perform a task through experience or the data learned. Any substance resulting in the body immune reaction is called as antigen. The immune reaction in the body is performed by secreting some proteins called as antibodies [16].
The natural immune system involves various levels. The first level prevents entering the outsider creatures or antigen through the skin. In the next level, the body is equipped with an innate immune system which generally copes with outsiders. The immune response at this level is the same against all antigens. The acquired immunity is the next level, with a customized coping method for any given antigen. Antigen is recognized by the white blood cells known as lymphocytes [17]. The algorithms designed for artificial immune system mainly model the acquired mechanism; apply in solving a wide range of computer problems. The artificial immune systems designed algorithms can be categorized into several groups: negative selection algorithm, Clonal selection algorithm, immune networks algorithm, and theory of danger [18 and 19]. The main idea of the Clonal selection method is to multiply only the cells whose antibodies are able to recognize the antigens [20, 21 and 22]. For negative selection algorithm, this idea is to produce a number of detectors and apply them for a new data categorization in the form of insider and outsider. In artificial immune systems, creation of a stable memory structure to tolerate antigens' further attacks is considered as the main idea [23 and 24]. In other words, the immune system ability to respond again to the same antigen may increase following by immune system reaction to a stranger, outsider antigen. The main difference between danger theory and the classic view is that in danger theory the human immune system does not respond to all insider cells, rather responds merely to those dangerous insiders [25].

### IV. RELATED WORKS

In [26] the balancing of load between the mesh routers is provided by using Optimized Link State Routing protocol (OLSR) with Expected Transmission count (ETX) i.e. OLSR-ETX. They modified the OLSR-ETX to prop up the wired-cum-wireless WMN. The modified new OLSR-ETX routing protocol is named as Wired-cum-Wireless WMN OLSRETX (W3-OLSR-ETX). Results show that W3-OLSR-ETX is better than AODV.

One of the key factors of the OLSR routing protocol is the MPR selection algorithm, which is based only on the reachability of each neighbor, not taking into consideration how they are moving. As a result, the selected MPR set may be unstable. One way to improve the stability of the MPR set is through spatial mobility metrics that are able to promptly monitor the degree of movement correlation between a node and its neighbors. Mr. cavalcanti showed that current metrics have limitations on capturing the spatial correlation in the various states of collective motion. Through an enhanced spatial mobility metric, they propose a MPR selection algorithm, which was integrated into a new mobility-aware OLSR protocol. they proposed a mobility-aware adaptive OLSR routing protocol, which is based on a new algorithm for MPR selection. The original MPR algorithm is based only on the number of reachable neighbors (a density metric) for defining the MPR set, not taking into account how nodes are moving. In contrast, the proposed solution adds a spatial mobility metric called Improved and Smoothed Degree of Spatial Dependence (ISDSD), so that the neighbors that have both a high reachability but also a high spatial movement correlation is selected. As a result, the selected MPR set tends to remain unchanged for a longer time, resulting in greater stability of the routes, which makes the protocol more efficient. The new technique provided a performance gain in terms of packet delivery ratio and end-to-end delay, besides presenting fewer out of order packets [27]. Chen et al, proposed a high-throughput routing protocol for wireless sensor networks through extending the OLSR protocol with opportunistic routing and network coding. Opportunistic routing and network coding leverages the receiver and transmitter diversity. Opportunistic routing is able to leverage the wireless channel's characteristic of broadcasting and opportunistically deliver data through multiple routing paths. In addition, OLSR can provide the information about network topologies and other parameters that opportunistic routing needs but cannot gain by itself. The results show that the proposed routing protocol can achieve much higher throughput than the OLSR protocol [28]. Ouacha et al. [29] described another link-based OLSR adaptation. The proposed method considers that nodes periodically exchange their positions, so that they can estimate the direction of motion and the remaining time that the node remains as a neighbor. The RWP model was the only employed in the modeling and evaluation of the proposed solution. Tamil selvi [30] proposed the secured OLSR protocol for MANET. The author presented the MPR selection based on BEST MPR selection, which reduced the number of TC message generated. Hence, the routing overhead is reduced in the network. Threshold cryptography was applied to the selected MPR nodes to provide security. The secret key of the source is split into number of shares based on count of MPR nodes in the network. The destination can pull through the TC message only if threshold numbers of shares are provided. The main disadvantage of this method was when threshold number of shares was compromised. This can be overcome by the share update method mechanism. This is proposed in the next section. In paper [31] they proposed new routing algorithm named Energy Saver Path Routing using Optimized Link State Routing (ESPR-OLSR) protocol because routing in MANET is serious issue because network topology which is changeable due to nodes mobility. Routing algorithm uses specific metrics to determine the optimum path between senders and receivers such as shortest minimum cost and minimum total power transmission etc. Many routing protocols have been proposed in last few years. Especially energy efficient routing is most important because all the nodes are limited battery power. Failure of one node may affect the entire networks. If a node runs out of energy, the probability of network partitioning will be increased. Since every mobile node has, limited power is become one of the main threats to the lifetime of the MANET. So routing in MANET should be in such a way that it will use the remaining battery power in an efficient way to increase the lifetime of the node network. Cervera et al. [32] presented Disjoint Multipath OLSR (DM−OLSR) function to address the following problems: 1) a partial view of the network topology, 2) flooding disruption attacks, and 3) load balancing in OLSR based networks. In DM−OLSR, the nodes select their MPRs with additional coverage during the topology discovery phase and compute, when possible, t+1 strictly disjoint paths during the route computation phase. To increase the chances of computing multiple disjoint paths from a source node to a destination node, during the topology discovery phase, the node select their MPR set with additional coverage and with the TCR parameter as zero. DM−OLSR improves the network topology view of the system nodes, and handles eventual flooding disruption attacks to the multipath construction mechanism. H¨arri et al. [33] defined the concept of Kinetic Multipoint Relaying (KMPR) where, instead of a node being periodically added to the MPR set, it is added for a period of time, which is estimated from the nodes' velocities. The authors evaluated the KMPR algorithm in scenarios generated by the RWP model. The adapted OLSR protocol showed a reduction in the number of broadcast retransmissions and end-to-end delay. The main limitations of that work are three: (1) assumption of constant velocity during the time the nodes remain neighbors; (2) disregarding the node pause time in modeling and evaluation of the algorithm; and (3) only the RWP was used. Mr Zhihao Guo and et al [34], presented Energy Aware OLSR ( OLSR_EA). Their Energy Aware OLSR labeled as OLSR_EA measures and predicts per-interval energy consumptions using the well-known Auto-Regressive Integrated Moving Average time series method. they develop a composite energy cost, by considering transmission power consumption and residual energy of each node, and use this composite energy index as the routing metric. OLSR-EA is able to prolong network lifetime and save total energy in MANET scenarios with a variety of traffic loads, node mobilities, and both homogeneous and heterogeneous power consumptions among the nodes. Cervera et al. [35] presented taxonomy of flooding disruption attacks that affect the topology map acquisition process in Hierarchical OLSR (HOLSR) network, and preventive mechanisms to mitigate the effect of this kind of attack. According to their work, it is possible to mitigate the effect of flooding disruption attack by selecting MPR set with additional coverage or generating control traffic with redundant information.

## V. PROPOSED ROUTING ALGORITHM: AIS-OLSR

Among the most important features in selecting a suitable route, one can name three parameters including route hop counts, remaining energy in the intermediate nodes, and the distance among nodes. Hop count is inversely related to route value; the higher is the hop count, the more probable is the route to be unsuitable. The remaining energy in the intermediate nodes is directly related to route value; the higher the route energy, it is wiser to take that route as once the intermediate nodes energy is depleted, the route will be dropped and transmission will be interrupted. Besides, selecting the routes with higher energy content leads to energy consumption unified distribution in the mobile ad hoc nodes, considered as a critical issue in mobile ad hoc networks constraining energy problem. The third parameter is the distance between source and destination nodes in the mobile ad hoc networks, which contributes finding the shortest route in terms of length between two source and destination nodes through a routing process. As previously mentioned on performance of OLSR protocol, to detect their neighbors, the nodes initially transmit a HELLO message to neighbors, store the delivered information in a table and distribute TC messages in the network using MPR points. Thus, all the networks nodes are aware about the existing connections and connection details to each node. The related information is stored in a table for each node.

### A. Composition of AIS-OLSR

As previously mentioned, a large number of algorithms have been purposed for artificial immune systems each of which applied in various domains. In the present work, negative selection and Clonal G algorithms were applied.

### B. Using Negative Selection Algorithm

Negative Selection algorithm creates based on T cells. T cells distinguish insider and outsider cells. It has two stages, the first one, which is learning stage, is like teamwork, and ends; it refers cells that identify and remove insiders. Then, stage two, which is test or implementation phase, compares antigens with remaining T cells of first stage, and removes if identified. The major function of this algorithm is identifying pattern.

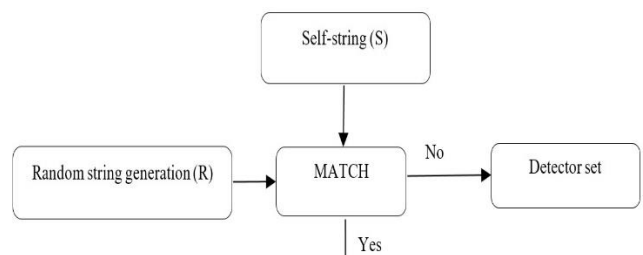

Fig. 1. Negative Selection algorithm learning.

In this regard, these algorithms are used to create a set of antibodies selecting the optimum route among them as follows:

| **Algorithm 2:** Negative Selection Algorithm |
|---|
| 1: **Input:** A  S ⊂ U ("self-set"); a set M ⊂ U ("monitor set"); an integer n |
| 2: **Output:** For each element m ∈ M, either "self" or "non-self" |
| 3: **Procedure**  Training phase<br>4: { <br>5:   d ← **empty set**<br>6:     while \|D\|< n do<br>7:   d ← **random detector**<br>8: } |

Fig. 2. Pseudo-code Negative Selection algorithm.

The source node in the standard OLSR through reviewing its routing table and the routes to the destination selects that route with minimum hop counts using the Dijkstra's algorithm. However, the process taken in the present work is as follows: The source node picks the routes, from routing table, leading to destination, but to select the optimum route, first, it applies the negative selection algorithm. In this algorithm, the antibodies are the routes reaching to destination in the routing table, while antigen is the mechanism, which tests two conditions including routes' energy and hop counts. Each time, through segregation phase, one antibody (route) is compared to one antigen until comparing all antibodies. Then, the worst routes in terms of energy and hop counts are rejected. During comparing antigen with antibodies (routes) being rejected or kept, each antibody (route) is compared to an antigen. If the given antibody (route) energy content is less than the threshold energy of the intermediate nodes, it is rejected; otherwise, it is entered to an array being analyzed in terms of hop counts.

| **Algorithm 1:** Pseudo-code comparing Antigen with Antibody |
|---|
| 1: **Input:** Antigen (Route's) |
| 2: **Output:** Array of Routes |
| 3: **Procedure**  Comparing Antigen with Antibody |
| 4: { |
| 5: **If**  energy(node $_i$) <  Threshold then |
| 6: { |
| 7:     Delete (Route $_i$) |
| 8: **Else  if** |
| 9: { |
| 10:   Array     ← Route $_i$ |
| 11:   Array Sort   Order by   hop count |
| 12: } |
| 13: } |
| 14: **If**  hop count (Route $_i$)  <  hop count ( Array Route)  then |
| 15: { |
| 16: Delete (Route $_i$) |
| 17: **Else if** |
| 18:   Max (hop count)  ←   Route |
| 19: } |
| 20: } |

Fig. 3. Pseudo-code comparing Antigen with Antibody.

This threshold is calculated by the formula 1,

$$\frac{The\ energy\ of\ node\ i}{Maximum\ initial\ energy\ of\ the\ nodes} \qquad (1)$$

Where in Equation (1), (i) is the intermediate nodes of each route. The number of arrays is decided based upon the number of antibodies (routes) intended to be in the group. Each route passing through the previous step enters to the array and the array is arranged based on the total hops until the destination. Then, by entering the next route, it is compared to the array. If route hop counts is greater than that of the routes in the array it will be rejected, otherwise it may replace a route with maximum hop count (and that route is rejected from the array) and the array is rearranged. This process is followed until all the routes are tested and those remained in the array enter to the detection set. Therefore, according to the negative selection algorithm, if the given antibody (route) matches with conditions (energy of the intermediate node is low and hop counts is high), the route will be rejected; otherwise it is shifted to the next phase – detection set. Indeed, instead of separating the insiders from outsides, the better routes are separated from the worse ones and the better ones are selected as the members of detection set.

In the next phase, two other actions are needed to be followed: 1) If necessary, hyper-mutation is performed; and 2) the best antibody (optimum route) is selected and kept in the immune memory, which is done using the Clonal G algorithm in this work.

### c. Using ClonalG algorithm

CLONALG algorithm, using its critical property, optimization, is introduced as the best approach in this area. The algorithm creates early cells, and selects colony on each antigen. Then, resulting antibodies will be used as initial memory cells in next iteration; the process retrieves until end condition, which is usually implementing determined replicas. Thereby, memory cells in each iteration can be created with higher affinity. Considering affinity plays a critical role in cells colonization. In fact, higher affinity causes greater proliferation and lower affinity will lead to less proliferation. On the other hand, mutation, which inversely relates with affinity, also plays a key role in this algorithm, namely higher affinity, less mutation.

---

**CLONAL-G Algorithm**

1. **Initialize:** Create a random population of individuals

2. **Antigenic Presentation:** For each antigenic pattern, do

   2.1. **Affinity Evaluation:** present antigen to each member of Population and determine affinity.

   2.2. **Clonal Selection and expansion:** Select n highest affinity Elements of population. Clone these with rates proportional to affinity.

   2.3. **Affinity maturation:** mutate all clones with rates inversely Proportional to affinity and add them to population.

   2.4. **Memory:** keep element of population with highest affinity to Antigen.

   2.5. **Meta-dynamics:** replace the m lowest affinity elements of Population with new ones.

3. **Cycle:** Repeat step 2 until stopping criterion is met.

---

Fig. 4. Pseudo-code CLONALG algorithm

TABLE I. CORRESPONDENCE BETWEEN IMMUNE SYSTEM AND CLONALG ALGORITHM

| Immune system | CLONAL-G |
|---|---|
| Antigen | Best routes in terms of energy and step |
| Antibody | Studying energy and step conditions |
| Affinity index | Proportion of total route nodes' energy to hop count |
| Mutation | Comparing routes in term of distance |

Antigens, provided at this stage, are the very antigen set of former stage superior in terms of energy and steps comparing other antigens. Antibodies structure also studies energy status and route steps.

### d. Affinity

Different studies refer antigen and antibody binding level as both distance and affinity [33]. The present research measures affinity by ratio of route nodes total energy to step numbers of all affinity routes; then, selects routes with the highest affinity. Therefore, routes with highest affinity will be selected and remained in later steps and other routes will be removed.

### e. Mutation and colonization

Once algorithm identified routes with higher affinity, mutation will initiate, if needed. Mutation rate depends on affinity, meaning that if affinity is high, no mutation takes place and security memory saves the route so that source node selects this route in sending packets to destination. On other side, routes' close affinity causes mutation. In fact, routes are initially ordered based on the highest affinity in a set; next, N number of this set with higher affinity will be selected to mutate. Mutation, here, compares routes in term of another criterion namely distance criterion, and selects that route with the shortest distance between source and destination. Finally, solution will be selected from remaining routes at the last step. The best route is the one with the most energy and least distance. This optimized route places in memory, which will be introduced as the best route for data transfer (Fig.5). AIS-OLSR protocol performance to OLSR and EAOLSR protocol, which is an improved version of OLSR protocol in term of energy level, is presented using packet delivery rate, end-to-end delay, network throughput, and network lifetime.

---

**For all Routes Calculate :**

   **Affinity**= (Energy Route Nodes) / (hopcount)

   **If** Affinity (Route $_i$) > Max Affinity then
      Self-Memory ← Route $_i$

   **Else**
      {
   **Mutation**

      **For** j=1 to N do
         {
         Distance (Route $_j$)

         Self-Memory ← Minimum (Route $_j$)
         }
      }

---

Fig. 5. Pseudo-code Mutation and colonization.

### f. implementation issues

As earlier stated, OLSR basic protocol operates with the shortest hop count and uses Dijkstra's algorithm for routing. It is assumed that all nodes are equipped with a geographic positioning system (GPS) always knowing their coordinates. Through applying the proposed method in OLSR algorithm, three new fields including "geographical position", "distance", and "energy" are added to the HELLO message packet. Here, the geographical position field is used to measure the distance between nodes, while the distance field is used to transfer the distance between nodes in any jump to the intermediate node. Finally, the energy field indicates the amount of remaining energy.

| Bits: | 0 1 2 3 4 5 6 7 8 9 0 1 2 3 4 5 6 7 8 9 0 1 2 3 4 5 6 7 8 9 0 1 | | |
|---|---|---|---|
| OLSR header: | Packet Length | | Packet Sequence Number |
| Message: | Message Type | V time | Message Size |
| | Originator Address | | |
| | Time To Live | Hop Count | Message Sequence Number |
| | MESSAG | | |
| Message: | Message Type | V time | Message Size |
| | Originator Address | | |
| | Time To Live | Hop Count | Message Sequence Number |
| | MESSAG | | |
| Message: | Message Type | V time | Message Size |
| | Originator Address | | |
| | **Energy** | **Distance** | **Geographic** |
| | MESSAG | | |

Fig. 6. New format of Message HELLO packet.

Each node starting to transmit HELLO message, first puts zero value in the distance field, longitude and latitude values in the geographical positioning field, and its energy content value in the energy field then send to the neighboring nodes. Based on the delivered longitude and latitude values, the receiving node in turn calculates the distance using eq. 2 and sums it up to the value in distance field and keeps it in its table as distance. Then, it transmits this value, its geographical position, and its energy content in response to node relaying HELLO message. Therefore, after the HELLO message is distributed, all nodes are having a table in which detecting all their neighbors; identifying their distance to neighboring node and the energy content of the neighboring nodes:

$$D = \sqrt{(x_1-x_2)^2 + (y_1-y_2)^2} \quad (2)$$

In Equation (2), (x1, y1) and (x2, y2) are the geographic positions of the node communicating the HELLO message, D is distance between source and destination node and the neighboring node, respectively. Then, each node sends its own and neighbors information in the form of a TC message including three distance, longitude and latitude, as well as energy fields, with hop count and number fields (which are in the main frame of the protocol) to the MPR points through which TC messages are distributed in the network. Once the TC messages are distributed, all network nodes will have a table consisting of all nodes information utilized in routing process. Through the standard OLSR protocol, only hop counts criterion is used for routing. However, in the method purposed in this work, two other criteria including energy and distance are also considered in the artificial immune system.

## VI. PERFORMANCE EVALUATION

To show performance of the AIS-OLSR routing protocol in comparing with the standard version of OLSR and EOLSR protocol That is an improved version of the OLSR protocol in terms of energy, , some criteria including packet delivery rate, end to end delay, throughput, and Network life time were applied .

Simulation was carried out in a NS2 (network simulator 2) [32] environment and the artificial immune system was implemented using the C++ programming language.

TABLE II. SIMULATION PARAMETERS

| Parameters | Value |
|---|---|
| Channel Type | Channel/Wireless channel |
| Publication Type | Two ray ground |
| Network Interface | Wireless Phy |
| Antenna | Omni Antenna |
| Simulation Area (m x m) | 1000 X 1000 |
| MAC layer | MAC/802.11 |
| Traffic Type | CBR |
| Queue Type | Drop Tail |
| Number of nodes | 100 |
| Primary energy | 10 Jules |
| Threshold | 0.5 Jules |
| Time simulation | 200 s |

### A. Packet delivery rate (PDR)

PDR equals the number of successfully delivered data packets delivered to destination nodes to the total number of transmitted data packets from the source node [37]. Thus, we can define PDR as shown in Equation (3).

$$PDR = \frac{Receive\ packets}{Sent\ packets} * 100 \quad (3)$$

As shown in fig. 7, the protocol presented in this work (AIS-OLSR) involves more desired PDR than that of OLSR and EA-OLSR, due to selecting better and more optimized routes.

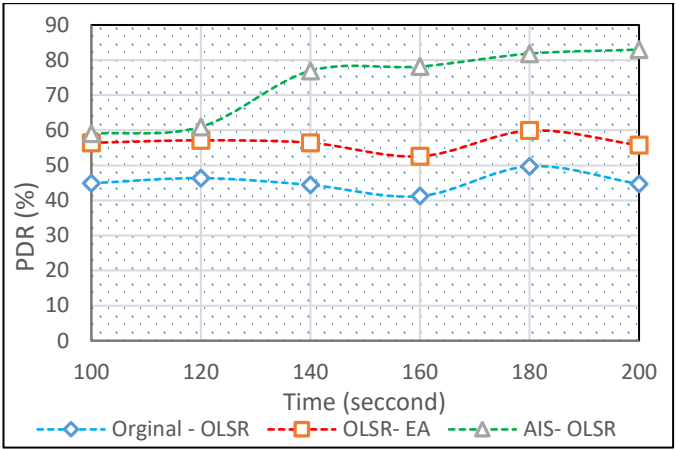

Fig. 7. PDR vs pause time.

### B. End to end delay

End to end delay sent by node (i) (source node) to packet j which is temporarily delivered to destination is as follows. Thus, we can define $End_{to end_{delay}}$ as shown in Equation (4).

$$End_{to end_{delay}} = Start_{time_{i,j}} - End\_time_{i,j} \qquad (4)$$

Where, $Start_{time_{i,j}}$ is the delivery time of packet j from node i and $End\_time_{i,j}$ is delivery time of this packet by destination node. As shown in fig. 8, the proposed protocol AIS-OLSR end-to-end delay is less than that of the standard OLSR protocol and EA-OLSR as selecting the optimum routes in terms of energy, hop count, and distance.

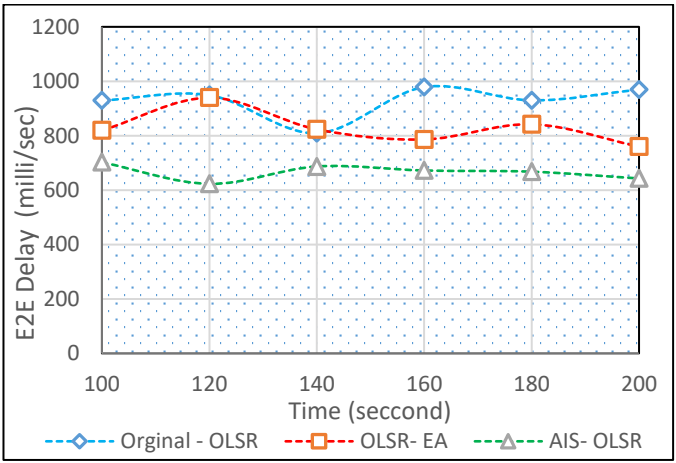

Fig. 8. End to End Delay

### C. Throughput

Throughput is regarded as the best criterion to compare the efficiency of routing protocols, obtained from dividing the destination delivered data to the data delivery time. Criteria such as PDR and end-to-end delay are also engaged in throughput: the more these criteria are, the higher the throughput would be. Fig.9 presents throughput in OLSR, EA-OLSR and AIS-OLSR protocols. This increase in throughput value in AIS-OLSR to OLSR and EA-OLSR is attributed to selecting better routes and the increased PDR is related to the reduction in end-to-end delay. AIS-OLSR protocol successfully delivered more amounts of data in a shorter time since the optimum routes had been selected.

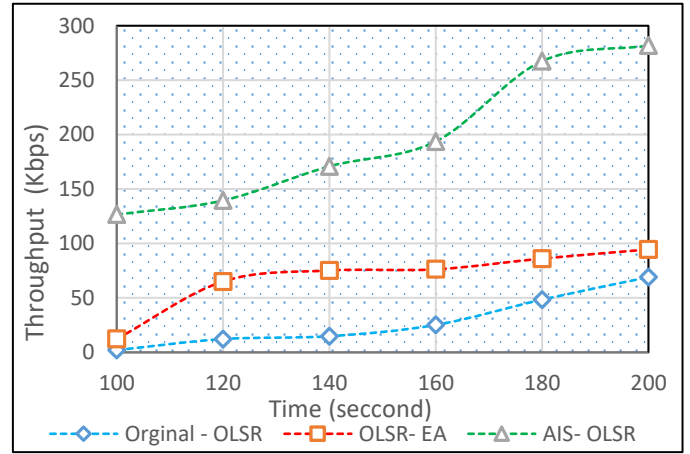

Fig. 9. Throughput vs pause time.

### D. Network Life Time

Node remaining energy is one of major issues in mobile ad-hoc networks presented here. As stated, consumed energy level directly influences network lifetime; therefore, network lifetime increases using high-energy routes. Fig. 10 shows that suggested protocol (AIS-OLSR) outperforms other two protocols in network lifetime indicating supremacy of this protocol in energy usage and increased network lifetime.

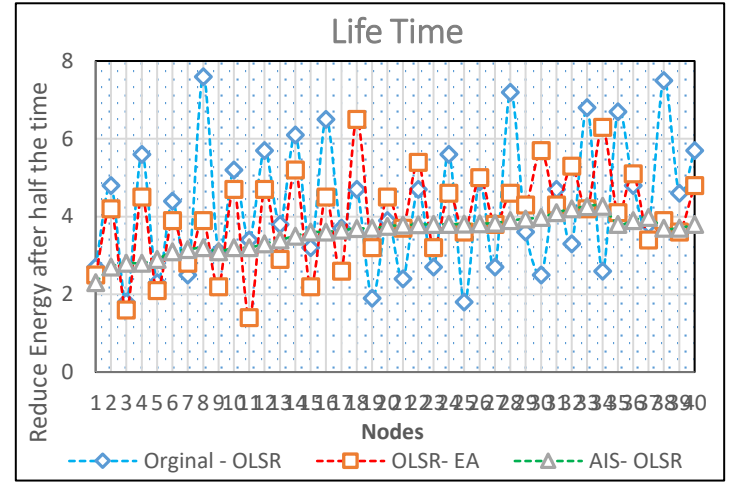

Fig. 10. Network life time vs ones.

### VII. CONCLUSION

In this paper, the OLSR protocol was applied to study selecting the optimum route among the available routes during mobile ad hoc networks routing process. Therefore, the artificial immune system was applied to select the best, optimum route. Three parameters including hop counts, intermediate nodes energy contents, and source and destination nodes distances were applied in this work to select the optimum route, whereas through the standard OLSR, only hop counts criterion is applied. The simulation results AIS-OLSR protocol indicated that artificial immune system could improve routing protocol efficiency in terms of end-to-end delay decrease, throughput increase, raising the number of delivered data packets and network lifetime increase.